\documentclass{article}
\usepackage[T1]{fontenc}
\usepackage{xcolor}
\pdftrailerid{}
\usepackage{microtype}
\usepackage{graphicx}
\usepackage{booktabs}
\usepackage{amsmath}
\usepackage{amssymb}
\usepackage{amsthm}
\usepackage{multirow}
\usepackage{makecell}
\usepackage{mathrsfs}
\usepackage{xspace}
\usepackage{xurl} %
\usepackage{hyperref}

\usepackage[accepted]{mlsys2025}
\makeatletter
\renewcommand{\printAffiliationsAndNotice}[1]{%
\stepcounter{@affiliationcounter}%
{\let\thefootnote\relax\footnotetext{\hspace*{-\footnotesep}\ifdefined\isaccepted #1\fi%
\forloop{@affilnum}{1}{\value{@affilnum} < \value{@affiliationcounter}}{
\textsuperscript{\arabic{@affilnum}}\ifcsname @affilname\the@affilnum\endcsname%
\csname @affilname\the@affilnum\endcsname%
\else
{\bf AUTHORERR: Missing \textbackslash{}mlsysaffiliation.}
\fi
}.
\ifdefined\mlsyscorrespondingauthor@text
Correspondence to: \mlsyscorrespondingauthor@text.
\else
{\bf AUTHORERR: Missing \textbackslash{}mlsyscorrespondingauthor.}
\fi
}
}
}
\makeatother
\usepackage[nameinlink,capitalise,noabbrev]{cleveref}
\crefformat{section}{#2\S#1#3}\Crefformat{section}{#2\S#1#3}
\crefformat{subsection}{#2\S#1#3}\Crefformat{subsection}{#2\S#1#3}
\crefformat{appendix}{#2Appendix~#1#3}\Crefformat{appendix}{#2Appendix~#1#3}
\crefformat{subappendix}{#2Appendix~#1#3}\Crefformat{subappendix}{#2Appendix~#1#3}
\crefmultiformat{appendix}{Appendices~#2#1#3}{ and~#2#1#3}{, #2#1#3}{, and~#2#1#3}
\crefrangeformat{appendix}{Appendices~#3#1#4--#5#2#6}

\theoremstyle{plain}

\newcommand{\method}{CoordRoute\xspace}
\newenvironment{compactitems}{%
  \setlength{\leftmargini}{1.1em}%
  \begin{itemize}\setlength{\itemsep}{1pt}\setlength{\parskip}{0pt}}%
  {\end{itemize}}

\mlsystitlerunning{CoordRoute: Coordinated Multi-Model Routing for Maximum Compute-Budget Utility}

\begin{document}

\twocolumn[
\mlsystitle{CoordRoute: Coordinated Multi-Model Routing for Maximum Compute-Budget Utility}

\begin{mlsysauthorlist}
\mlsysauthor{Boning Huangfu}{sjtu}
\mlsysauthor{Boxiao Du}{sjtu}
\mlsysauthor{Chen Chen}{sjtu}
\end{mlsysauthorlist}
\mlsysaffiliation{sjtu}{Shanghai Jiao Tong University, Shanghai, China}
\mlsyscorrespondingauthor{Chen Chen}{chen-chen@sjtu.edu.cn}
\mlsyskeywords{LLM routing, model serving, resource allocation, Model-as-a-Service}

\vskip 0.3in

\begin{abstract}
Model routing sends each request to one of several large language models (LLMs) of different sizes.
In practice, requests share a fixed compute budget: a spending cap on rented model APIs or the capacity of a private GPU cluster.
Most existing routers do not account for the total budget: each decides a request on its own, so per-request optima are not optimal overall.
We advocate \emph{coordinated routing}, which allocates the budget across requests to maximize its utility, their total satisfaction, along a \emph{predict--plan--rectify} route: predict what each model would give and cost, plan the choice against later requests, and rectify it during generation.
Each step faces a challenge: a request's gain from each model is uncertain before generation, later arrivals are unknown when a request must start, and observing answers during generation consumes the budget itself.
\method addresses them with relative demand prediction, which compares candidate models through past requests that lead to the same routing choices; a lookahead solver, which charges each choice for what its compute could earn for later requests; and budget-aware runtime correction, which drops or upgrades in-flight requests at the same price, inspecting answers only when it pays off.
On Azure Conversation and BurstGPT traces, \method improves total satisfaction by 18.72\% over the strongest existing router.
\end{abstract}
]

\printAffiliationsAndNotice{}

\section{Introduction}
\label{sec:introduction}

Model-as-a-Service (MaaS) has become a standard way to deliver large language models (LLMs): several models of different sizes are offered behind one interface.
Since a small model already answers many requests well while others need a much stronger one, choosing a model for each request, known as \emph{model routing}, is the main lever for serving requests well at low cost.
Existing routers~\citep{ong2025routellmlearningroutellms,chen2023frugalgptuselargelanguage,hu2024routerbench,somerstep2025carrotcostawarerate} optimize each request on its own, trading its answer quality against its own cost, with no bound on what all requests spend together.

In practice, total spending is bounded: serving runs under a \emph{fixed compute budget} (\cref{tab:budget-vs-no-budget}).
An organization that rents model APIs caps its total spending~\citep{litellm2026budgets}.
Besides, a private-cloud provider, which keeps an organization's data in-house, serves all its models from one fixed GPU cluster.
Either way, a call is not a separate bill but a share of a fixed total, while load changes within minutes.
A stronger model for one request leaves less compute for the requests that follow, so the right model for a request depends on how many others compete: a strong model is worthwhile under light load and wasteful under heavy load.
The goal of routing thus changes: under a fixed budget, it must maximize the total utility, that is, the sum of user satisfaction over all requests, and choices that are optimal for each request are not optimal for this sum.

Most routers are \emph{non-coordinated}: each decides a request on its own, regardless of what other requests need (\cref{sec:background:related-work}).
Under a fixed budget, this is costly: a trade-off tuned for one load fails at others.
CARROT~\citep{somerstep2025carrotcostawarerate}, for example, fixes its quality--cost weight in advance.
Under the same budget and a weight tuned for 100 requests per window, it leaves 41.7\% of the requests unserved when a window has 200 requests in total, and 57.6\% of the budget unused when it has 50 in total (\cref{fig:carrot-load}).
A few routers coordinate requests, but in a simplified form: they cover only a set fixed in advance, such as a queued batch, a known number of queries, or an assumed workload~\citep{mei2025omnirouter,zhang2024treacle,panda2025pilot,kasnavieh2026routerwise}, and cannot correct their plan during generation, so they misallocate compute once actual arrivals or outcomes deviate from it.

We advocate \emph{full-fledged coordinated routing}: a router should choose each request's model by weighing its gain against the \emph{opportunity cost} of the compute it takes, that is, what this compute would earn for the other requests, including those queued and those yet to arrive.
It follows a \emph{predict--plan--rectify} route: before a request starts, the router predicts what each candidate would give and cost and plans the choice against the opportunity cost; during generation, it rectifies the choice as the answer reveals its quality and length.
Each step raises one challenge (\cref{sec:motivation:challenges}):
\begin{compactitems}
  \item \textbf{C1: Uncertain per-request gain.} A request's quality and compute on each model are unknown before generation and vary widely even for one model, and running candidates to find out spends the compute being allocated. It is thus difficult to predict accurately at low cost.
  \item \textbf{C2: Unknown future load.} The opportunity cost depends on requests that have not arrived and whose number changes within minutes, yet each request must be routed when it starts. The router must thus plan without knowing the future load.
  \item \textbf{C3: Costly runtime observation.} Rectifying requires judging a partial answer, which runs a model and, done often, can cost more than the generation itself. Frequent checks thus waste the budget, while rare checks come too late.
\end{compactitems}

We present \method, a router that coordinates requests to maximize the utility of a fixed compute budget, with one key idea per challenge:
\begin{compactitems}
  \item \textbf{Relative demand prediction (\cref{sec:solution:prediction}).}
  Routing needs only how candidates compare, not their absolute quality.
  \method learns a request distance from routing consequences, so that requests are close when they favor the same models, and reads scores from past neighbors.
  This achieves accurate and efficient prediction, at under 0.01\% of the generation compute.
  \item \textbf{Lookahead coordinated solving (\cref{sec:solution:solving}).}
  After one choice, the rest of the window is the same problem with a smaller budget.
  \method thus prices the budget left for unknown later requests by a benefit curve of recent requests, scaled by the expected arrivals, and re-solves before every decision.
  This plans every choice with later requests in view, in 0.92~ms per decision.
  \item \textbf{Budget-aware runtime correction (\cref{sec:solution:correction}).}
  Coordination continues during generation: \method drops or upgrades an in-flight request when, at the solver's price, the change raises the window's total satisfaction, and pays for an inspection only after enough generation to justify it.
  This rectifies wrong choices in time and at low cost.
\end{compactitems}

We implement \method as a Python library of about 3,500 LoC and evaluate it on request streams following the Azure Conversation and BurstGPT traces, with all routing, monitoring, and generation charged to one budget.
\method leads eight existing routers at all 21 budgets, improving total satisfaction by 18.72\% over the strongest, CARROT.
It completes 91.88\% of arrivals versus 80.13\% and reaches CARROT's best satisfaction with 43\% less computation.
It also performs well across all our sensitivity analyses.

\section{Background}
\label{sec:background}

\subsection{Routing under a Fixed Compute Budget}
\label{sec:background:services}

\begin{figure}[t]
  \centering
  \includegraphics[width=.92\columnwidth]{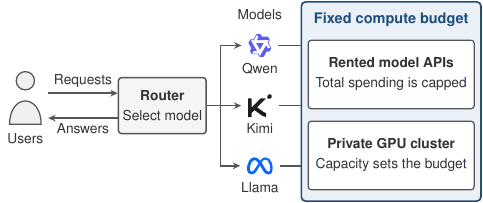}
  \caption{Model serving under a fixed compute budget, set by a spending cap on rented APIs or by the capacity of a private GPU cluster. Logos are illustrative; \cref{sec:experiments:setup} lists the evaluated models.}
  \label{fig:background-architecture}
\end{figure}

\paragraph{Fixed compute budgets.}
LLMs come in families of several sizes, such as six dense Qwen3 models from 0.6B to 32B parameters~\citep{yang2025qwen3technicalreport}; larger models answer better but need more computation.
Organizations serve such models under a fixed compute budget in two main forms (\cref{fig:background-architecture}).
In the first, an organization rents hosted models through APIs under a fixed total budget, which LLM gateways enforce across a proxy, team, user, or key~\citep{litellm2026budgets}.
In the second, capacity sets the budget: a private-cloud MaaS provider serves the models from its own fixed GPU cluster.
Serving systems can multiplex several models on shared GPUs~\citep{li2023alpaserve,duan2024muxserveflexiblespatialtemporalmultiplexing} and shift capacity among them, but the total is fixed.
In both forms, load fluctuates sharply within minutes (\cref{fig:arrival-uncertainty}), and the compute for each request is committed when it starts, before later requests are known.

\paragraph{Satisfaction and utility.}
Users care whether an answer meets their request; we call how well it does so the answer's \emph{satisfaction}, on a scale from 0 to 1.
A judge model, DeepSeek-V4.1-Flash (\cref{app:setup}), scores each anonymized answer against the request's requirements; strong LLM judges agree with human preferences about as often as humans agree with each other~\citep{zheng2023judgingllmasajudgemtbenchchatbot}.
The \emph{utility} of a compute budget is the total satisfaction it yields over all arriving requests, which the provider seeks to maximize; a request left without an effective answer contributes zero.

\paragraph{Routing without and with a total budget.}
\Cref{tab:budget-vs-no-budget} contrasts the two problems.
Existing routers~\citep{ong2025routellmlearningroutellms,chen2023frugalgptuselargelanguage,hu2024routerbench,ding2024hybridllm,aggarwal2024automix,chen2024routerdc} leave the total unbounded: requests are treated as independent, so a per-request quality--cost trade-off suffices, and optimizing each request also optimizes the whole.
A total budget changes the problem: every call draws on one budget, requests compete, the load determines which model each should get, and a poor choice costs other requests their answers rather than costing extra money.
A local optimum for each request is then no longer the global optimum.

\begin{table}[t]
\centering
\small
\setlength{\tabcolsep}{5pt}
\begin{tabular}{@{}lll@{}}
\toprule
 & No total budget & Fixed total budget \\
\midrule
Cost of a call & Separate bill & Share of one budget \\
Other requests & Independent & Compete for compute \\
Load & Does not matter & Decides the right model \\
A poor choice & Overpays once & Starves other requests \\
Local optimum & Also global & Not global \\
\bottomrule
\end{tabular}
\caption{Routing without and with a total budget.}
\label{tab:budget-vs-no-budget}
\end{table}

\subsection{Existing Routers Are Not Fully Coordinated}
\label{sec:background:related-work}

\paragraph{Non-coordinated routing.}
We call a router \emph{non-coordinated} if its decision for a request depends only on that request and on a trade-off fixed in advance, not on what other requests need.
Most routers are non-coordinated, whether they route by a threshold, a predicted quality--cost score, or a cascade that calls a stronger model after a weak answer~\citep{ong2025routellmlearningroutellms,hu2024routerbench,somerstep2025carrotcostawarerate,chen2023frugalgptuselargelanguage,dekoninck2025unifiedapproachroutingcascading,wang-etal-2025-mixllm}.

\begin{figure}[t]
  \centering
  \includegraphics[width=.85\columnwidth]{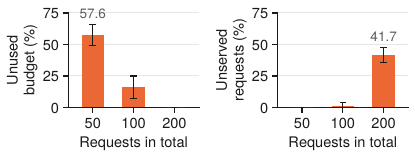}
  \caption{CARROT with a weight tuned for 100 requests per window under one budget. With fewer requests in total, compute goes unused (left); with more, requests receive no complete answer (right). Bars show means over 30 request sequences; whiskers show standard deviations.}
  \label{fig:carrot-load}
\end{figure}

Ideally, a router adapts whenever the combination of budget and requests changes, so that overall utility stays maximal; a fixed trade-off cannot.
Consider CARROT~\citep{somerstep2025carrotcostawarerate}, which predicts each model's quality and cost for a request and picks the best weighted score.
We take satisfaction as its quality and FLOPs as its cost, calibrate its weight for 100 requests sharing 4~PFLOPs over 300~s, and keep this budget while the window holds 50, 100, or 200 requests in total (\cref{fig:carrot-load}).
With 200 requests in total, CARROT spends as if requests were few: the budget runs out early, and 41.7\% of the requests receive no complete answer.
With 50 requests in total, it spends as if requests were many: 57.6\% of the budget stays unused, although this compute could have served these requests better.

\paragraph{Partially coordinated routing.}
Beyond non-coordinated routing, a few routers attempt to coordinate requests, but their coordination is heavily simplified and its effect limited; we call them \emph{partially coordinated}.
First, they simplify the \emph{dynamics of arrivals}, covering only a set of requests fixed when they decide.
OmniRouter jointly assigns the queries of one batch~\citep{mei2025omnirouter}, yet a batch holds only the requests queued at that moment: they are balanced against each other but not against those that arrive later, so an early batch can spend what a later burst needs.
TREACLE and PILOT spread one user's budget over a known number of queries~\citep{zhang2024treacle,panda2025pilot}; RouterWise plans routing shares for an assumed workload~\citep{kasnavieh2026routerwise}; and RouteBalance accounts only for requests already in service, through each instance's current latency~\citep{da2026routebalance}.
In realistic serving scenarios, however, most competing requests have not yet arrived, so an allocation planned for the queued, known, or assumed set over- or under-spends once actual arrivals deviate from it.
Second, they simplify the \emph{uncertainty of satisfaction}.
A request's actual quality and cost are settled only during real execution, but these methods plan on prior estimates; the decision stays fixed at dispatch or, in cascades, is revised only after the answer is complete~\citep{chen2023frugalgptuselargelanguage,dekoninck2025unifiedapproachroutingcascading}, when its compute is already spent.
Without online correction, every misprediction is paid in full, which limits their effect.

\section{Coordinated Routing}
\label{sec:motivation}

\subsection{Our Insight: Full-Fledged Coordinated Routing}
\label{sec:motivation:insight}

Our insight is that coordination must be \emph{full-fledged}, covering every competing request and every stage of spending.
It spans all requests that share the budget, whether queued, yet to arrive, or still being generated; it continues from before generation to the end of generation; and it counts all compute, including that of routing and monitoring.
One price of compute links these decisions: each request's model is chosen by the request's own gain minus the \emph{opportunity cost} of the compute it takes, valued by what that compute would earn for the other requests.
Compute thus flows to where it raises total satisfaction most.

To model this, we divide service time into \emph{allocation windows} of a few minutes: the requests arriving in a window share its \emph{compute budget} $B$, the inference computation available to them in floating-point operations (FLOPs) across all models.
This covers both settings of \cref{sec:background:services}: an API user divides its total budget among windows, and a private cluster with an effective inference rate of $P$ FLOP/s gives $B=PT$ for a window of $T$ seconds.
In our model, budget left idle in a window is not saved for a later one, and routing depends on the budget per request, not on the absolute budget.
FLOPs measure what routing controls, the work each call adds~\citep{kaplan2020scaling}, regardless of how the serving system batches, schedules, or places the call; those choices set $P$.
Our gains do not depend on exact FLOP counts (\cref{sec:experiments:sensitivity}).

\paragraph{Problem formulation.}
Let $\mathcal{M}$ be the model pool and $\mathcal{R}$ the requests arriving within a window.
For request $i$ and model $m$, let $q_{i,m}\in[0,1]$ be the satisfaction of the resulting answer and $c_{i,m}$ its computation in FLOPs, covering input processing and output generation, and let the binary variable $x_{i,m}$ indicate whether request $i$ is assigned to model $m$.
The goal is to maximize the budget's utility, the total satisfaction of the window's requests:
\begin{equation}
\label{eq:routing-objective}
\begin{aligned}
\max_x\quad
&\sum_{i\in\mathcal{R}}\sum_{m\in\mathcal{M}}q_{i,m}x_{i,m} \\
\text{s.t.}\quad
&\sum_{i\in\mathcal{R}}\sum_{m\in\mathcal{M}}c_{i,m}x_{i,m}\leq B, \\
&\textstyle\sum_{m\in\mathcal{M}}x_{i,m}\leq 1\ \ \forall i,
\quad x_{i,m}\in\{0,1\}\ \ \forall i,m.
\end{aligned}
\end{equation}
The constraints bound total computation and assign at most one model per request; an unserved request contributes zero.
The shared budget couples all choices: whether one request deserves a stronger model depends on the gains and costs of all the others.

\subsection{Predict--Plan--Rectify and Its Challenges}
\label{sec:motivation:challenges}

A coordinated router cannot solve \cref{eq:routing-objective} in advance, because requests arrive and are answered over time.
It instead follows a \emph{predict--plan--rectify} route for each request: it \emph{predicts} each candidate's satisfaction and computation before the request starts, \emph{plans} the choice against what the remaining budget can earn for later requests, and \emph{rectifies} the choice during generation.
Each step lacks information that \cref{eq:routing-objective} assumes, which raises one challenge.

\begin{figure}[t]
  \centering
  \includegraphics[width=.72\columnwidth]{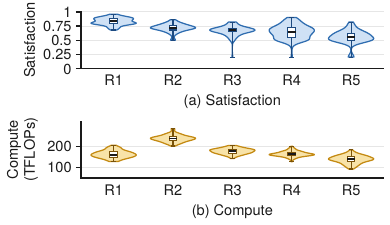}
  \caption{Five requests (R1--R5), each with 100 Qwen3-32B generations under fixed settings; computation uses a log axis.}
  \label{fig:outcome-compute-uncertainty}
\end{figure}

\paragraph{C1 in predicting: uncertain per-request gain.}
Predicting requires the satisfaction $q_{i,m}$ and computation $c_{i,m}$ of each candidate, which remain unknown even after request $i$ arrives.
Computation depends on the output length~\citep{kwon2023pagedattention}, and with sampling even a fixed request and model can produce answers with different meanings~\citep{farquhar2024semanticentropy}.
Across 100 Qwen3-32B answers to each of five UltraFeedback~\citep{cui2023ultrafeedback} requests under fixed settings, both quantities vary widely (\cref{fig:outcome-compute-uncertainty}; \cref{app:setup}); for request R4, satisfaction ranges from 0.20 to 0.90 and computation from 119.5 to 199.4~TFLOPs.
Measuring them directly defeats the purpose, because running candidates first spends the compute that routing allocates.
The predictor must therefore be accurate without running the candidates.

\begin{figure}[t]
  \centering
  \includegraphics[width=\columnwidth]{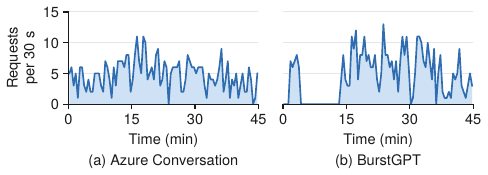}
  \caption{Arrivals in 30-second bins over 45 minutes, for the $2\times$-load replay streams of the evaluation (seed 0).}
  \label{fig:arrival-uncertainty}
\end{figure}

\paragraph{C2 in planning: unknown future load.}
Planning weighs request $i$ against all other requests in the window, but when $i$ must start, later requests have not yet arrived.
The provider knows neither how many will come nor which models they will need, and arrivals fluctuate sharply over short intervals in both the Azure Conversation~\citep{azure2023llminferencetrace} and BurstGPT~\citep{wang2025burstgpt} traces (\cref{fig:arrival-uncertainty}).
Underestimating later load spends the budget too early and leaves later requests unserved, while overestimating it holds back current answers and leaves the budget idle.
The router must thus commit compute before it knows what that compute is worth later.

\begin{figure}[t]
  \centering
  \includegraphics[width=\columnwidth]{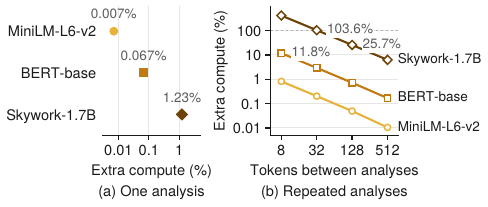}
  \caption{Computation of semantic analysis relative to full Qwen3-32B generation on 300 requests, for (a) one analysis at 128 generated tokens and (b) repeated analyses.}
  \label{fig:correction:monitorcost}
\end{figure}

\paragraph{C3 in rectifying: costly runtime observation.}
Rectifying relies on generation, which gradually reveals an answer's actual quality and length and could expose a wrong allocation before all its compute is spent.
Judging an unfinished answer, however, requires semantic analysis of the request and the generated prefix, and each analysis runs a model forward pass.
On 300 Qwen3-32B answers (\cref{fig:correction:monitorcost}), one BERT-base~\citep{devlin2019bert} analysis costs only 0.067\% of the generation, but repeating it every eight tokens adds 11.8\%, and running a Skywork-Reward-V2-Qwen3-1.7B evaluator~\citep{liu2025skyworkrewardv2} every 32 tokens adds 103.6\%, more than generating the answer itself.
Observation thus competes for the very compute it is meant to save, so it must be used sparingly.

\section{\method Design}
\label{sec:solution}

\begin{figure*}[t]
  \centering
  \includegraphics[width=0.75\textwidth]{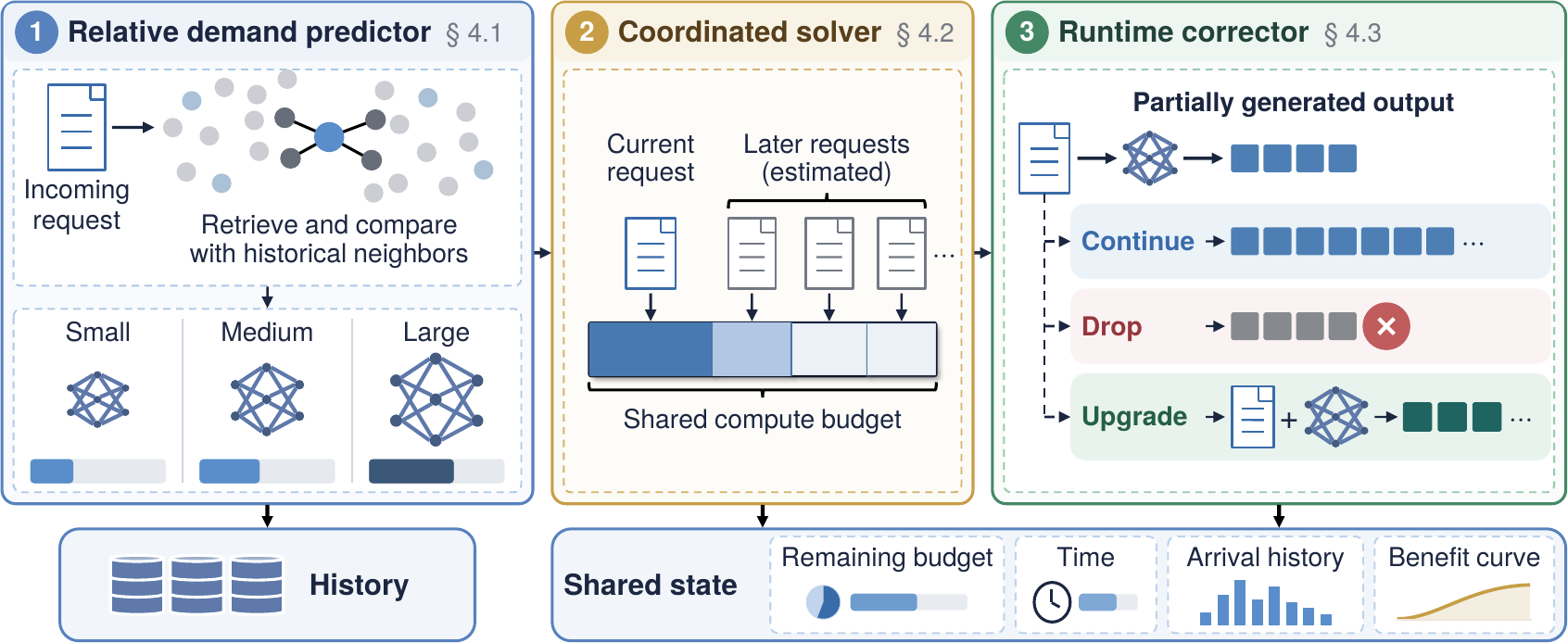}
  \caption{\method overview. (1) Predict: historical neighbors provide relative demand scores. (2) Plan: the solver divides the window budget between current and later requests. (3) Rectify: correction continues, drops, or restarts the full original request on a stronger model. Model sizes and allocations are schematic.}
  \label{fig:system-overview}
\end{figure*}

\method implements the three steps of coordinated routing with one component each (\cref{fig:system-overview}).
To predict, the \emph{relative demand predictor} scores each candidate model from past requests before the request starts (C1, \cref{sec:solution:prediction}); to plan, the \emph{coordinated solver} picks the model whose score plus the value of the budget it leaves for later requests is highest (C2, \cref{sec:solution:solving}); and to rectify, the \emph{runtime corrector} continues, drops, or upgrades the request during generation at the solver's price, inspecting only when it pays off (C3, \cref{sec:solution:correction}).
All three charge their computation to the window budget and share its state, so every decision pursues the window's total satisfaction.

\subsection{Relative Demand Prediction}
\label{sec:solution:prediction}

The solver needs to know how much each candidate would help a request before any answer exists, at a cost small enough to pay for every request.
Lightweight routers regress each model's absolute satisfaction from request features~\citep{hu2024routerbench,wang-etal-2025-mixllm}.
This target is a poor fit: it is hard, as outcomes vary widely even for one request and model (\cref{fig:outcome-compute-uncertainty}), and it is more than routing needs, because a choice depends only on how candidates compare.
On 47.0\% of our evaluation requests, Qwen3-32B and Kimi-K2.5 differ in satisfaction by at most 0.05, so a per-model loss such as mean squared error spends effort on errors that never change the choice.

\paragraph{Key idea.}
\method predicts \emph{relative demand scores}: one score per candidate on the satisfaction scale, learned to rank candidates correctly rather than to match absolute values.
It reads these scores from similar past requests, where similarity is defined by \emph{routing consequences}: two requests are close when they favor the same models, not when their texts look alike.
One lightweight encoding and a neighbor lookup thus replace running candidate LLMs.

\paragraph{Learning request similarity.}
Offline, each training request $i$ has recorded satisfaction $q_{i,m}$ and computation $c_{i,m}$ on every model $m\in\mathcal M$.
We describe it with $G$ feature groups $\mathbf z_i^{(g)}$ from a frozen text encoder, character n-grams, and its length, structure, task keywords, and input computation.
The distance between two requests weights these groups by $\theta_g\geq0$:
\begin{equation}
\label{eq:request-distance}
D_\theta(i,j)
=\sum_{g=1}^{G}\theta_g
\left\|\mathbf z_i^{(g)}-\mathbf z_j^{(g)}\right\|_2^2.
\end{equation}
To learn $\theta$, we measure what request $i$ loses by borrowing the outcomes of request $j$.
Under a computation penalty $\lambda$, a router that sees $j$'s satisfaction but pays $i$'s computation chooses model $a$ and loses $r_{i\leftarrow j}$ against the best choice for $i$:
\begin{equation}
\label{eq:borrowed-selection-loss}
\begin{aligned}
a&=\arg\max_{m\in\mathcal M}
\left(q_{j,m}-\lambda c_{i,m}\right),\\
r_{i\leftarrow j}(\lambda)
&=\max_{m\in\mathcal M}\left(q_{i,m}-\lambda c_{i,m}\right)
-\left(q_{i,a}-\lambda c_{i,a}\right).
\end{aligned}
\end{equation}
We average this loss over several penalties and fit $\theta$ by regularized nonnegative regression on nearby training pairs (\cref{app:predictor}), so requests whose outcomes lead to the same choices end up close.

\begin{figure}[t]
  \centering
  \includegraphics[width=.8\columnwidth]{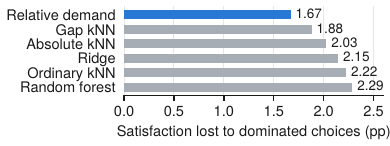}
  \caption{Satisfaction lost to dominated choices by six predictors with one shared feature set and a fixed solver, without correction; means over the three analysis scenarios of \cref{sec:experiments:setup} and 21 budgets (\cref{app:predictor}).}
  \label{fig:predictor-loss}
\end{figure}

\paragraph{Online prediction.}
For an arriving request $i$, the predictor retrieves its $k$ nearest training requests $\mathcal N_i$ under $D_\theta$ and averages their satisfaction with inverse-power distance weights $w_{ij}$:
\begin{equation}
\label{eq:relative-demand}
d_{i,m}=\sum_{j\in\mathcal N_i}w_{ij}q_{j,m},
\qquad
\sum_{j\in\mathcal N_i}w_{ij}=1.
\end{equation}
A length regressor and the model architecture give the computation estimate $\widehat c_{i,m}$.
In a controlled comparison with shared features and a fixed solver (\cref{fig:predictor-loss}), relative demand scores lose 11.3\% less satisfaction to \emph{dominated} choices, where another candidate answers better with no more computation, than the best of five alternatives, and 22.2\% less than ridge regression.

\subsection{Lookahead Coordinated Solving}
\label{sec:solution:solving}

The solver chooses a model when a request starts service.
It knows the request's scores and computation estimates, the window's remaining budget and time, and the arrivals so far, but not the requests still to come.
Non-coordinated routers price compute once, offline, for an assumed load, so they idle or exhaust the budget whenever the actual load differs (\cref{fig:carrot-load}).
Online pacing and knapsack thresholds adjust the price step by step as budget is spent~\citep{balseiro2023dualmirror,zhou2008onlineknapsack,panda2025pilot}, so a single large expense or a sudden burst moves the price only gradually.

\begin{figure}[t]
  \centering
  \includegraphics[width=.82\columnwidth]{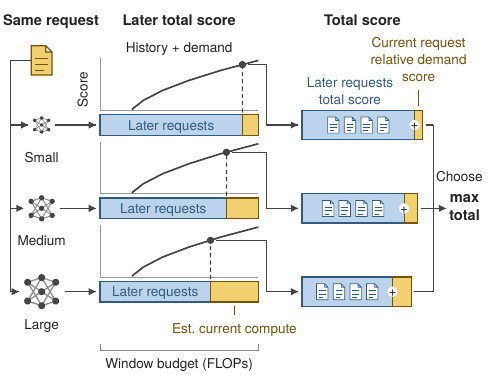}
  \caption{Coordinated model selection. A candidate's estimated computation (yellow) leaves a budget for later requests (blue), which the benefit curve maps to their total score. The solver adds the current relative demand score and chooses the highest total.}
  \label{fig:coordinated-solver}
\end{figure}

\paragraph{Key idea.}
Two properties of the problem let the solver anticipate later requests at each decision.
First, the budget belongs to one allocation window and does not carry over, so a decision only needs to plan for the rest of the window.
Second, the problem is self-similar: after a model is chosen, the rest of the window poses the same problem with a smaller budget.
A choice is therefore worth its own score plus what the remaining budget can still earn for later requests (\cref{fig:coordinated-solver}).
The drop in this value is the choice's opportunity cost, as in bid-price control, online knapsack, and online linear programming~\citep{talluri1998bidprice,zhou2008onlineknapsack,agrawal2014onlinelp}.
\method estimates it from recent requests and recomputes it before every decision~\citep{jasin2012resolving,bumpensanti2020resolving}, so one large expense is charged for its full effect on later requests, and a burst that fills the queue raises the price at once.

\paragraph{Benefit curve.}
Later requests are unknown, but recent requests are a good sample of them.
The solver therefore keeps the scores and computation estimates of the latest $h=128$ routed requests, $\mathcal H_t$, and asks what average score they could reach if each used $b$ FLOPs on average.
The answer is the \emph{benefit curve} $F_t(b)$; with $\alpha_{j,m}$ the fraction of request $j$ planned on model $m$,
\begin{equation}
\label{eq:benefit-frontier}
\begin{aligned}
F_t(b)=\max_{\alpha\geq0}\quad
&\frac{1}{h}\sum_{j\in\mathcal H_t}\sum_{m\in\mathcal M}
d_{j,m}\alpha_{j,m}\\
\text{s.t.}\quad
&\frac{1}{h}\sum_{j\in\mathcal H_t}\sum_{m\in\mathcal M}
\widehat c_{j,m}\alpha_{j,m}\leq b,\\
&\sum_{m\in\mathcal M}\alpha_{j,m}=1,
\quad \forall j\in\mathcal H_t.
\end{aligned}
\end{equation}
Because only the average is constrained, the curve gives more computation to the requests that gain more from it.
$F_t$ is piecewise linear, and sorting the penalties at which each request's preferred model changes builds it in $O(h|\mathcal M|\log(h|\mathcal M|))$ time.

The curve describes one average request; the remaining budget is shared by all requests still to come.
The solver therefore estimates their number $\widehat N_t$: the queued requests plus the arrivals expected in the rest of the window, estimated from the arrival history and an arrival prior, with a small margin against spending too early (\cref{app:solver}).
If budget $b$ is left, each of these requests receives $b/\widehat N_t$ on average, so together they can earn
$\widehat V_t(b)=\widehat N_t F_t(b/\widehat N_t)$.
This value is zero when no request remains and $-\infty$ when $b$ cannot cover the cheapest candidate of each remaining request.

\paragraph{Lookahead choice.}
Let $B_t$ be the window's remaining budget when request $i$ starts at time $t$.
Choosing model $m$ leaves $B_t-\widehat c_{i,m}$ for later requests, so the solver picks
\begin{equation}
\label{eq:recursive-pairing}
m_i^*
=\arg\max_{\substack{m\in\mathcal M\\\widehat c_{i,m}\leq B_t}}
\left[
d_{i,m}
+\widehat V_t\!\left(B_t-\widehat c_{i,m}\right)
\right]
\end{equation}
and moves on to the next request with the updated budget; if the budget cannot cover the expected load, it falls back to the cheapest model that can start.
In a controlled study over 15 windows of about 100 arrivals with shared scores, this choice beats a fixed per-request allowance by 24.9\% at 3~PFLOPs per window and reaches 90\% of an offline oracle that knows the full sequence and actual outcomes (\cref{app:solver}).

\subsection{Budget-Aware Runtime Correction}
\label{sec:solution:correction}

Generation reveals whether the initial choice still looks right: an answer may grow longer than expected or show signs that it will not meet the request.
Cascades use such evidence only after the answer is complete, when its computation is already spent~\citep{chen2023frugalgptuselargelanguage,dekoninck2025unifiedapproachroutingcascading}, and their gates judge each answer in isolation from the budget.
Monitoring during generation, in turn, can cost more than it saves (\cref{fig:correction:monitorcost}).

\paragraph{Key idea.}
\method extends coordination into generation.
It re-decides an in-flight request at the solver's price, so a correction happens only when the compute it frees or consumes is worth more to the window than the change in this request's satisfaction.
To keep this affordable, it treats each inspection as an expense under the same budget and pays for one only when enough generation has accumulated to justify it.

\paragraph{Pre-monitoring.}
Inspections may happen at geometrically spaced positions: 8, 16, 32, \ldots, 2048 generated tokens.
At each position, a low-cost \emph{pre-monitor} compares two quantities: $\Delta C$, the generation computation since the last full inspection, and $C_{\mathrm{check}}$, the estimated cost of the drop and upgrade checks available there.
It admits a full inspection only if $\Delta C\geq\gamma C_{\mathrm{check}}$, with $\gamma=4$, so inspections cost at most a quarter of the generation they oversee.
Both quantities follow from token counts and model architectures, without any forward pass.

\begin{figure}[t]
  \centering
  \includegraphics[width=.9\columnwidth]{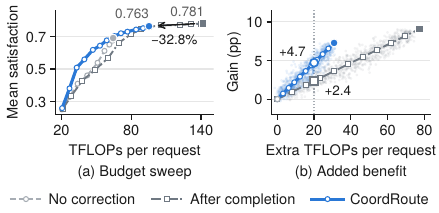}
  \caption{Correction when all 300 requests start on Qwen3-32B with Kimi-K2.5 as the target. (a) Satisfaction versus computation under per-request limits. (b) Gain versus extra computation as more requests become eligible (\cref{app:correction}).}
  \label{fig:correction-effect}
\end{figure}

\paragraph{Correction decisions.}
An admitted inspection encodes the prefix with a small text encoder, and drop checks also run the 1.7B evaluator over the request and prefix.
Lightweight heads trained on past generation traces then update the expected satisfaction and remaining computation of the current answer.
Both decisions value computation with the solver's benefit curve.
A drop gives up the request's satisfaction, so \method drops only when the computation saved is worth more to later requests, by a margin that accounts for prediction uncertainty.
An upgrade stops the current model and sends the complete original request, without the prefix, to a stronger model that the remaining budget can start.
\method upgrades only when the lower confidence bound of the satisfaction gain, predicted from paired past outcomes, is nonnegative and the window's total satisfaction is expected to rise by a margin (\cref{app:correction}).
In a controlled test where all requests start on Qwen3-32B with Kimi-K2.5 as the upgrade target, correction raises satisfaction by 4.7 percentage points for 20 extra TFLOPs per request, double the gain of upgrading after completion (\cref{fig:correction-effect}).

\section{Implementation and Discussion}
\label{sec:implementation}

\subsection{Implementation}
\label{sec:implementation:code}

We implement \method as a Python library of about 3,500 LoC.
The predictor uses a frozen MiniLM-L6-v2 encoder~\citep{wang2020minilm,reimers2019sbert} and 700 in-memory training requests, the solver runs on the CPU, and runtime inspection runs MiniLM-L6-v2 and a Skywork-Reward-V2-Qwen3-1.7B evaluator~\citep{liu2025skyworkrewardv2} on a GPU.

Each window has a ledger that charges every call its FLOPs at the actual input and output lengths; routing, pre-monitoring, inspection, and restarts are charged like generation.
For evaluation, we record answers and judged satisfaction for every request--model pair once and replay streams against these records, as routing benchmarks do~\citep{hu2024routerbench}, so all methods face the same requests, answers, and arrivals, and see labels only for training requests.

\subsection{Discussion}
\label{sec:discussion}

\paragraph{Compute metric.}
FLOPs are determined only by the model architecture and the number of tokens, and are the standard measure of model computation~\citep{kaplan2020scaling,hoffmann2022training}.
\method is therefore not tied to a particular model, hardware, or inference backend: differences in efficiency between environments enter mainly through the effective rate $P$.
It also extends to priced settings, with each call's price as $c_{i,m}$ and the spending cap as $B$.
We use compute because prices fluctuate widely and are affected by discounts and promotions, whereas FLOPs capture the stable trend.

\paragraph{Latency SLOs.}
\method is orthogonal to SLO optimization at the system level: serving systems can improve latency through scheduling and parallelism without changing the compute consumed~\citep{agrawal2024sarathi,zhong2024distserve,patel2024splitwise}.
It also helps them by relieving backlogs: queued requests count toward the expected number of remaining requests (\cref{sec:solution:solving}), so once a backlog forms, the opportunity cost of compute rises at once and requests move to smaller, faster models, an established way to protect latency under load spikes~\citep{zhang2020modelswitching,ahmad2024proteus,ahmad2025diffserve}.

\paragraph{Fixed GPU pools and redeployment.}
On a fixed GPU pool, each model's compute use follows the routing decisions, so GPUs must be redeployed among models as requests change.
Fast redeployment makes this a minor concern: existing systems scale model instances within seconds and let several models share GPUs~\citep{zhang2025blitzscale,yu2026prism,xiang2025aegaeon}, far faster than one window.
Our evaluation therefore treats redeployment as instantaneous: each window has $B=PT$, and requests are served in arrival order at rate $P$ (\cref{sec:experiments:setup}); since both settings reduce to \cref{eq:routing-objective} with a budget $B$, its conclusions also apply to API spending caps.

\section{Experiments}
\label{sec:experiments}

This section evaluates \method under fixed compute budgets.
After the setup (\cref{sec:experiments:setup}), we compare it with existing routers end to end (\cref{sec:experiments:end-to-end}), examine the effectiveness of each component (\cref{sec:experiments:ablation}), and measure its overhead (\cref{sec:experiments:overhead}) and robustness (\cref{sec:experiments:sensitivity}).

\subsection{Setup}
\label{sec:experiments:setup}

\paragraph{Data and models.}
We select 1,000 instruction requests from UltraFeedback~\citep{cui2023ultrafeedback}, drawn from ShareGPT, FLAN, UltraChat, and Evol-Instruct, and split them into 700 training and 300 evaluation requests, keeping near-duplicates in one split.
The candidates are Alpaca-7B~\citep{taori2023alpaca}, Qwen3-32B~\citep{yang2025qwen3technicalreport}, and Kimi-K2.5~\citep{kimiteam2026kimik25} (\cref{app:setup}).

\paragraph{Request streams.}
We assign evaluation requests, in repeated shuffled passes, to the timestamps of 45-minute intervals from the Azure Conversation~\citep{azure2023llminferencetrace} and BurstGPT~\citep{wang2025burstgpt} traces.
We thin both traces to $0.5\times$, $1\times$, and $2\times$ a reference load of 180 requests per stream.
Ten seeds per trace and load yield 60 streams with mean arrival counts of 103--451 across loads, 13--26\% above nominal and unknown to all methods.
Other analyses use three \emph{analysis scenarios} with ten seeds each: stationary Poisson and low- and high-load Azure streams.

\paragraph{Evaluation.}
Each stream is divided into nine 5-minute allocation windows that each receive one ninth of its budget, requests are served first come, first served on a clock advanced by charged FLOPs at rate $P$, and all routing, generation, monitoring, and restart FLOPs are charged.
The metric is mean satisfaction per arrival; unstarted, dropped, and truncated requests score zero.

\paragraph{Baselines.}
We compare against eight baselines: RouterBench kNN and MLP~\citep{hu2024routerbench}, RouteLLM~\citep{ong2025routellmlearningroutellms}, CARROT~\citep{somerstep2025carrotcostawarerate}, FrugalGPT~\citep{chen2023frugalgptuselargelanguage}, Unified Routing and Unified Cascade~\citep{dekoninck2025unifiedapproachroutingcascading}, and MixLLM~\citep{wang-etal-2025-mixllm}.
For each budget, we calibrate their quality--cost parameters or policy searches on training data so that spending fits the budget at the reference load, the only load an operator knows before traffic arrives (\cref{app:setup}).

\subsection{End-to-End Performance}
\label{sec:experiments:end-to-end}

\begin{figure}[t]
  \centering
  \includegraphics[width=\columnwidth]{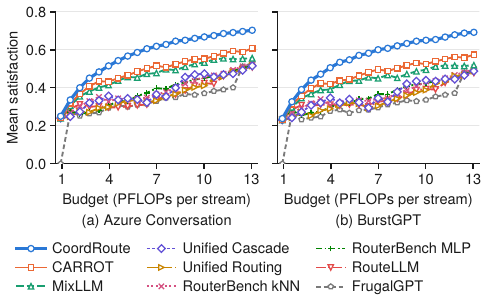}
  \caption{End-to-end satisfaction across 21 budgets; each curve averages three loads with ten seeds each.}
  \label{fig:end-to-end}
\end{figure}

\begin{figure}[t]
  \centering
  \includegraphics[width=\columnwidth]{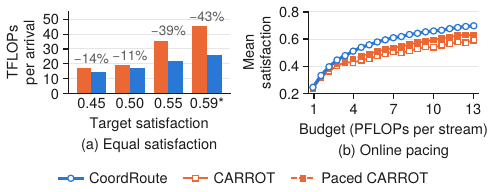}
  \caption{Against CARROT on the two traces. (a) Compute per arrival of the cheapest budget run that reaches each satisfaction; * marks CARROT's best. (b) CARROT with its penalty paced online by dual mirror descent.}
  \label{fig:strong-baselines}
\end{figure}

\begin{figure}[t]
  \centering
  \includegraphics[width=\columnwidth]{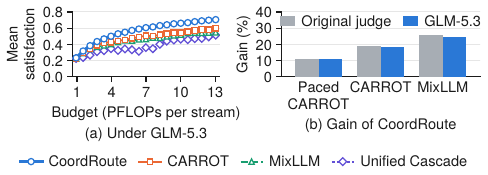}
  \caption{Second judge. (a) Two-trace satisfaction rescored by GLM-5.3. (b) Average gain of \method over the strongest baselines under each judge.}
  \label{fig:second-judge}
\end{figure}

We sweep 21 budgets from 0.92 to 13.05~PFLOPs per stream, from 3\% of the computation needed to answer every arrival with Kimi-K2.5 at the heaviest load to 155\% at the lightest.
\method leads at all 21 budgets on both traces (\cref{fig:end-to-end}); averaged over budgets and traces, it reaches 0.5672 versus 0.4778 for CARROT, the strongest existing baseline, an 18.72\% gain (95\% CI 15.48--22.26\%; \cref{app:setup}).
This gain comes partly from serving more requests with the same budget: the solver keeps budget for later arrivals, so \method completes 91.88\% of arrivals versus 80.13\%, and only 2.92\% cannot start, against 18.92\% for CARROT, whose fixed trade-off exhausts busy windows early (\cref{app:end-to-end}).
Its choices also follow congestion, as \cref{sec:discussion} anticipates: within the same runs, it sends 30.3\% of requests to Kimi-K2.5 when more than ten requests wait, against 57.0\% when none wait, whereas CARROT sends 45.0\% and 45.7\%.
\method also chooses better models for the requests it serves.
With our online pacing by dual mirror descent~\citep{balseiro2023dualmirror}, tuned in hindsight, CARROT completes nearly as many requests, 90.39\%, yet \method stays 10.89\% higher and leads at every budget; it reaches paced CARROT's best satisfaction with a 34\% smaller budget (\cref{fig:strong-baselines}b; \cref{app:end-to-end}).
At equal satisfaction, \method reaches CARROT's best with 43\% less computation (\cref{fig:strong-baselines}a; \cref{app:end-to-end}).
Under GLM-5.3, a judge from another model family, \method still leads every baseline at all 21 budgets with nearly unchanged gains (\cref{fig:second-judge}; \cref{app:end-to-end}).

\subsection{Ablation Studies}
\label{sec:experiments:ablation}

\begin{figure}[t]
  \centering
  \includegraphics[width=\columnwidth]{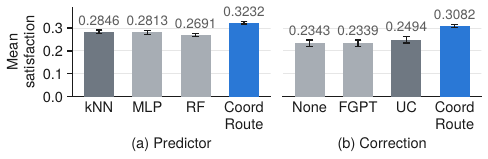}
  \caption{Ablations of (a) the predictor on 13 code-related requests, against RouterBench kNN/MLP and MixLLM RF heads, and (b) correction on 34 long-answer Evol-Instruct requests, against FrugalGPT (FGPT) and Unified Cascade (UC) gates. Means over 20 Poisson streams with 95\% bootstrap intervals.}
  \label{fig:ablation}
\end{figure}

\begin{figure}[t]
  \centering
  \includegraphics[width=\columnwidth]{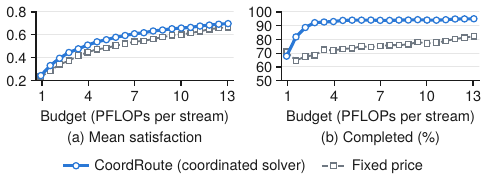}
  \caption{Coordinated solver ablation on the two traces: \method against the same system with a fixed price per FLOP calibrated like the baselines, both with runtime correction.}
  \label{fig:solver-ablation}
\end{figure}

\paragraph{Effectiveness of the relative demand predictor.}
The predictor matters most when the stronger candidates differ.
On 13 code-related requests, with a mean gap of 0.266 between Kimi-K2.5 and Qwen3-32B (0.187 overall), we swap in the absolute-satisfaction heads of RouterBench and MixLLM.
Ours scores 0.3232, 13.54\% above the best replacement, RouterBench kNN, at similar computation (\cref{fig:ablation}a; \cref{app:ablation}).

\paragraph{Effectiveness of the coordinated solver.}
We replace the solver with a fixed price per FLOP, calibrated like the baselines to spend 1/180 of the stream budget per request on training forecasts.
With this price, our predictor and corrector already lift satisfaction 7.07\% above CARROT on average, to the level of paced CARROT.
The coordinated solver adds another 10.88\% and leads at all 21 budgets, mainly by completing more requests (91.88\% versus 74.83\%) with similar computation (\cref{fig:solver-ablation}).
Without correction, it also leads two allocators with the same information, PILOT's online-knapsack threshold~\citep{panda2025pilot} and dual mirror descent pacing with the best of nine step sizes, at all 21 budgets by at least 11.04\% and 3.15\% (\cref{app:ablation}).

\paragraph{Effectiveness of runtime correction.}
On the two traces, correction raises satisfaction at 20 of 21 budgets, by 3.72\% on average and up to 6.54\% at intermediate budgets (\cref{app:ablation}).
It matters most for long answers: on Poisson streams of the 34 Evol-Instruct requests, full correction is 23.57\% above the strongest gate, Unified Cascade's, and 31.54\% above no correction (\cref{fig:ablation}b; \cref{app:ablation}).

\subsection{Overhead}
\label{sec:experiments:overhead}

\begin{figure}[t]
  \centering
  \includegraphics[width=\columnwidth]{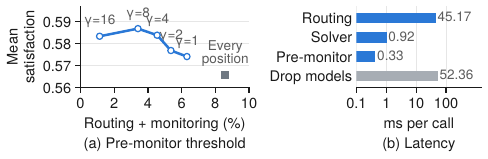}
  \caption{Overhead. (a) Satisfaction versus routing and monitoring share on the analysis scenarios, for thresholds $\gamma$ and inspection at every position. (b) Mean latency per call on one CPU thread; drop models run on a GPU.}
  \label{fig:overhead}
\end{figure}

On one CPU thread, routing takes 45.17~ms, including 0.92~ms for the solver, and a pre-monitor check 0.33~ms (\cref{fig:overhead}b).
On the three analysis scenarios, pre-monitoring with $\gamma=4$ cuts full inspections by 53.0\% and monitoring computation by 48.4\% relative to inspecting at every position, and raises satisfaction from 0.5658 to 0.5839 (\cref{app:overhead}).
Every $\gamma$ from 1 to 16 is cheaper and better than inspecting at every position (\cref{fig:overhead}a); the deployed $\gamma=4$, set in development, is within 0.5\% of the best and keeps the overhead at 4.55\% of all computation.

\subsection{Sensitivity and Robustness}
\label{sec:experiments:sensitivity}

\begin{figure}[t]
  \centering
  \includegraphics[width=\columnwidth]{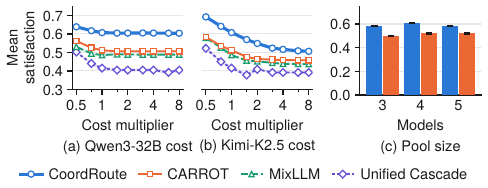}
  \caption{Sensitivity on the analysis scenarios to (a, b) a model's cost over three budgets and (c) pool size; 95\% bootstrap CIs.}
  \label{fig:sensitivity}
\end{figure}

\begin{figure}[t]
  \centering
  \includegraphics[width=\columnwidth]{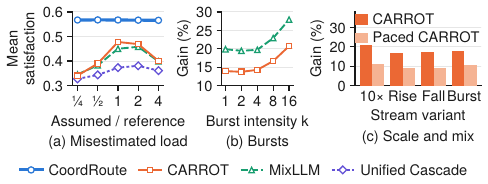}
  \caption{Robustness. (a) Assumed loads of 0.25--4$\times$ the reference. (b) Gain of \method with bursts at $k\times$ the background rate. (c) Gain at $10\times$ scale and with drifting task mixes.}
  \label{fig:robustness}
\end{figure}

\paragraph{Gains hold when relative model costs change.}
As memory-bound decoding or large mixture-of-experts weights can make actual cost differ from FLOPs, we multiply the generation cost of Qwen3-32B or Kimi-K2.5 by 0.5 to 8 in prediction, charging, and baseline calibration.
\method improves on CARROT by 10.58--19.82\% averaged over three budgets and leads CARROT, MixLLM, and Unified Cascade at each budget (\cref{fig:sensitivity}a, b; \cref{app:robustness}).

\paragraph{Gains hold with more models and at larger scale.}
With Qwen2.5-0.5B-Instruct~\citep{qwen2025qwen25} and then Qwen3.8-2.4T-A95B added and predictors and policies refitted, \method is 16.44\% and 12.35\% above CARROT with four and five models (\cref{fig:sensitivity}c) and leads all baselines at every budget (\cref{app:robustness}).
With ten times the arrivals and budget, \method improves on CARROT and paced CARROT by 20.81\% and 11.29\%, leading at all budgets (\cref{fig:robustness}c; \cref{app:robustness}).

\paragraph{A misjudged load barely affects \method.}
If the reference load is misjudged by 0.25--4$\times$ and all methods use the wrong value, \method, which updates its arrival estimate online, changes by at most 0.25\% on average, while CARROT and MixLLM, recalibrated to it, lose up to 28.6\% and 23.1\% (\cref{fig:robustness}a).
\method leads every baseline under every assumption, and stays ahead on average even when they are calibrated to the actual mean load of each trace and load level (\cref{app:robustness}).

\paragraph{Gains grow with burstiness and survive task-mix drift.}
With synthetic bursts at $k$ times the background rate, the gain over CARROT and MixLLM rises from 13.95\% and 19.89\% at $k=1$ to 20.73\% and 27.91\% at $k=16$ (\cref{fig:robustness}b).
When expensive requests concentrate early, late, or in bursts, \method changes by under 2\% and leads CARROT by 16.91--17.96\% and its paced variant by 9.27--10.64\% (\cref{fig:robustness}c).
Solver settings change satisfaction by at most 0.4\% on average (\cref{app:robustness}).

\section{Conclusion}
\label{sec:conclusion}

Under a fixed compute budget, routing must maximize total utility rather than each request's own trade-off.
\method realizes full-fledged coordinated routing along a predict--plan--rectify route, improving satisfaction by 18.72\% over the strongest existing router.
As it only decides how the budget is shared, it runs on top of existing serving systems.

\bibliographystyle{mlsys2025}
\bibliography{main}

\section*{AI Usage Disclosure}
We used two generative AI assistants, GPT-6.1 Sol (OpenAI, through Codex) and Claude Opus 5.5 (Anthropic, through Claude Code).
The authors defined the problem, designed \method, and planned the experiments; the assistants played a supporting role, discussing design alternatives and helping write and debug code for experiments and figures.
In literature search, they suggested related work; the authors checked every cited paper and its bibliographic entry.
In writing, they edited and revised the authors' drafts for language and clarity.
The authors reviewed all AI-assisted code, results, and text, and take full responsibility for the paper.
Separately, language models are objects of study in our experiments: they serve as candidate models, as satisfaction judges (DeepSeek-V4.1-Flash and GLM-5.3), and as components of \method (MiniLM-L6-v2 and Skywork-Reward-V2-Qwen3-1.7B), as described in \cref{sec:background,sec:implementation,sec:experiments}.

\clearpage
\appendix
\crefalias{section}{appendix}\crefalias{subsection}{subappendix}
\section{Evaluation Setup}
\label{app:setup}

\paragraph{Models and computation.}
Alpaca-7B runs locally, and the answers of Qwen3-32B and Kimi-K2.5 are obtained once through their providers' APIs.
Each answer is charged the FLOPs implied by its model's architecture and its actual input and output token counts, about twice the active parameters per token plus attention~\citep{kaplan2020scaling}; routing and monitoring are charged in the same units.

\paragraph{Satisfaction judge.}
DeepSeek-V4.1-Flash scores each answer in two stages.
Before seeing any answer, it reads only the request and writes a fixed rubric: the expected useful outcome, explicit requirements, critical correctness checks, and the appropriate response mode (answer, clarify, or refuse).
It then scores each answer on its own against this rubric, without the answering model's identity, price, or other answers.
Tasks with a determinate outcome score 1 if correct and 0 otherwise; other tasks are graded against request-specific anchors at 0, 0.2, 0.5, 0.8, and 1, with scores between anchors allowed, where a central technical error normally caps the score at 0.2.
Empty answers score 0, and answers that reach the generation limit are graded as supplied.

\paragraph{Request outcomes.}
Each arrival completes, never starts because the window budget cannot cover routing or the selected model, is truncated when the window budget runs out, or is dropped by the corrector.
Unstarted, truncated, and dropped requests score zero, so every reported mean covers all arrivals.

\paragraph{Repeated generations.}
\cref{fig:outcome-compute-uncertainty} uses five requests (a project plan, a logistics plan, an authentication policy, a training proposal, and an ensemble plan), with mean satisfaction 0.832, 0.715, 0.668, 0.647, and 0.554.
Qwen3-32B answers each 100 times with a fixed prompt, thinking disabled, a 16,384-token limit, and default sampling; none of the 500 answers is discarded.

\paragraph{Baselines and budget calibration.}
RouterBench, CARROT, RouteLLM, and MixLLM take a quality--cost parameter rather than a budget.
For each budget, we apply each candidate parameter with the method's own rule to out-of-fold predictions on the 700 training requests and select the one with the largest mean computation per request, routing included, that stays within 1/180 of the stream budget, matching the reference load of 180 arrivals, which is set from the Azure arrival rate of the preceding ten minutes; the parameter then stays fixed, as in the original methods.
Candidates are the 32 RouterBench willingness-to-pay values plus its change points, zero and 121 log-spaced CARROT penalties from $10^{-6}$ to $10^6$, the same values for MixLLM, and RouteLLM thresholds for strong-model shares of $0,0.05,\ldots,1$.
RouteLLM keeps its two-model design with Alpaca-7B and Kimi-K2.5; MixLLM keeps fixed predictors, because its online updates need feedback that no method observes, and drops its latency-oriented waiting-time penalty.
FrugalGPT and Unified Routing and Cascade run their own training-time policy searches for each budget.
RouterBench's predictions and choices match its official implementation.

\paragraph{Partially coordinated methods.}
TREACLE and OmniRouter need a known number of queries or a known batch, which would give them information no other method has.
RouterWise serves with fixed prices from an offline plan for an assumed workload, which the calibrated baselines and the fixed-price ablation represent.
RouteBalance adds current instance latency, which the budget objective excludes.
PILOT spreads a budget over a known number of queries with a threshold that depends only on the budget already spent; \cref{app:ablation} evaluates this threshold as an allocator, with the expected number of remaining requests in place of the known count.

\paragraph{Uncertainty.}
For end-to-end gains, 10,000 paired bootstrap replicates resample the 300 evaluation requests and the ten seeds within each trace and load.
The 95\% interval is 15.48--22.26\% for the 18.72\% gain over CARROT and 8.03--14.04\% for the 10.89\% gain over paced CARROT; no replicate shows a gain at or below zero.
Studies on fixed request pools resample traffic seeds only.

\section{Relative Demand Predictor}
\label{app:predictor}

\paragraph{Configuration.}
Each request has 384 MiniLM-L6-v2 features, 23 request features, and up to 12,000 character TF--IDF features of 3--5-grams, in seven distance groups: the embedding, the TF--IDF features, and five groups of request features.
A training pair enters the borrowed-selection loss when either request is among the other's 100 nearest neighbors before metric learning, and near-duplicate groups never borrow from each other.
Nonnegative fitting uses regularization strength 1, and prediction averages 40 neighbors with weights proportional to $D_\theta(i,j)^{-2}$.
Historical scores come from three group-disjoint out-of-fold partitions of the 700 training requests; the 300 evaluation requests are never used for fitting.

\paragraph{Routing diagnostic.}
\cref{fig:predictor-loss} disables runtime correction and keeps the three analysis scenarios, ten seeds, 21 budgets, and nine windows.
The relative predictor and centered-gap kNN share 407 features, five folds, 15 neighbors, and inverse-distance weights, and differ only in the metric-learning target; the other predictors use the same features (\cref{tab:routing-diagnostic}).
For an arrival completed on model $a$, the dominated-choice loss is $\ell_i=\max_{m:c_{i,m}\leq c_{i,a}} q_{i,m}-q_{i,a}$, and $\ell_i=0$ otherwise.
As this loss does not penalize unfinished requests or cheap low-quality choices, we also checked satisfaction and completion, on which the relative predictor is also highest; the 95\% interval of its loss reduction over gap kNN is 6.97--14.94\%.
The deployed predictor, with its own feature set, also has a lower loss than all five alternatives.

\begin{table}[t]
\centering
\small
\setlength{\tabcolsep}{6pt}
\begin{tabular}{lr}
\toprule
Predictor & Loss (pp) \\
\midrule
Relative, shared & 1.6682 \\
Gap kNN & 1.8800 \\
Absolute kNN & 2.0269 \\
Ridge & 2.1452 \\
Ordinary kNN & 2.2235 \\
Random forest & 2.2862 \\
\bottomrule
\end{tabular}
\caption{Route-only diagnostic on the analysis scenarios: satisfaction lost to dominated choices (lower is better).}
\label{tab:routing-diagnostic}
\end{table}

\section{Coordinated Solver}
\label{app:solver}

\paragraph{Arrival estimate.}
When a request is routed, the solver estimates the number of requests still to be served in the window as
\[
\widehat N_t=Q_t+\mu_t+0.75\sqrt{\mu_t+\mu_t^2/\kappa_t},\qquad
\mu_t=\frac{\kappa_t\,\tau_t}{5/r_0+t},
\]
where $Q_t$ counts the other queued requests, $\tau_t$ is the time left in the window, $t$ the time since the stream began, $\kappa_t=5+n_t$ with $n_t$ the arrivals so far, and $r_0$ the reference rate of 180 arrivals per 2,700~s.
$\mu_t$ is the expected number of remaining arrivals under a Poisson rate with a Gamma prior worth five arrivals, and the margin adds 0.75 standard deviations of the resulting negative-binomial count.
The same $\widehat N_t$ scales the benefit curve in solving and correction.

\begin{figure}[t]
  \centering
  \includegraphics[width=.9\columnwidth]{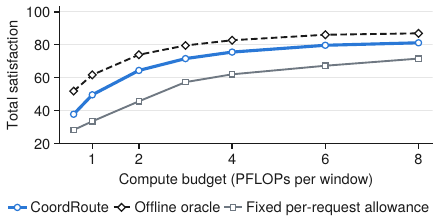}
  \caption{Total satisfaction per window over 15 windows of 84--124 arrivals, without correction; online methods share the same scores, and the oracle knows actual outcomes.}
  \label{fig:solver-online-allocation}
\end{figure}

\paragraph{Controlled allocation study.}
\Cref{fig:solver-online-allocation} compares the solver with a fixed allowance of $B/100$ per request and an offline oracle over 15 windows of 84--124 arrivals.
The online methods share fixed scores and the known computation of the current request, but never see future requests or actual satisfaction; the oracle knows the full sequence and actual outcomes.
At 3~PFLOPs per window, total satisfaction is 71.65 for the solver, 57.38 for the fixed allowance, and 79.62 for the oracle.

\section{Runtime Correction}
\label{app:correction}

\paragraph{Inspection positions and cost.}
The pre-monitor costs $2\times10^6$ FLOPs plus 2,000 FLOPs per byte of request and prefix, a median of 3.4~MFLOPs.
A drop check costs a median of 444~GFLOPs, 99.6\% of it in the 1.7B evaluator, and an upgrade check 6.1~GFLOPs.
A check is available where at least 20 training answers of the source model are still running: drop checks up to 256 tokens for Alpaca-7B and 2,048 for the other models, and upgrade checks up to 256 tokens for Alpaca-7B and 2,048 for Qwen3-32B.

\paragraph{Outcome distributions.}
Each check represents the possible outcomes of the current answer as weighted samples $s$ with weights $w_s$, built from the 700 training requests with group-disjoint folds.
For drops, a ridge head predicts the answer's satisfaction and log remaining length from prefix features (TF--IDF and MiniLM-L6-v2 embeddings, text statistics, and the evaluator's score and hidden state), and cross-fitted residuals at the same model and position form the samples.
For upgrades, the samples are training requests whose answer from the same source model is still running at this position, paired with their answers from stronger models and shifted toward the current request by LightGBM heads~\citep{ke2017lightgbm} for gain, remaining work, and restart cost.
In both cases, 90\% of the weight goes to the 40 most similar training answers and 10\% is spread uniformly.

\paragraph{Decision rules.}
Let $B_t$ be the free window budget, $r_s$ and $q_s$ the remaining computation and satisfaction of the source answer in sample $s$ ($q_s=0$ if it would be truncated), and $\widehat V_t(b)=\widehat N_t F_t(b/\widehat N_t)$.
For a weighted quantity $x$, let $\mathrm{LB}(x)=\bar x-\big[\sum_s w_s(x_s-\bar x)^2\cdot\sum_s w_s^2\big]^{1/2}$ with $\bar x=\sum_s w_sx_s$.
A drop is recommended when
\[
\mathrm{LB}\big(\widehat V_t(B_t)-\widehat V_t(B_t-r_s)-q_s\big)>0.02,
\]
that is, when the computation the answer would still use is worth more to later requests than its own satisfaction; a second test repeats this comparison with window satisfaction normalized by that of fixed models at the same computation per arrival.
A recommended drop first triggers an upgrade check, and the request is dropped only if no upgrade is accepted.
An upgrade with restart cost $c_s$ and satisfaction $t_s$, both capped by the free budget, is accepted when
\[
\mathrm{LB}(t_s-q_s)\geq0,\qquad
\mathrm{LB}\big(W(c_s,t_s)-W(r_s,q_s)\big)>0.02,
\]
where $W(c,q)$ is the normalized window satisfaction when this request scores $q$ and later requests receive $\widehat V_t(B_t-c)$, with the benefit curve compressed by 1.5 along its computation axis.
The restarted answer runs to completion without further checks.
The thresholds, the pre-monitor ratio $\gamma=4$, and the factor 1.5 were carried over from development on an earlier request set and not tuned on this paper's data.

\paragraph{Local correction effect.}
\cref{fig:correction-effect} starts all 300 evaluation requests on Qwen3-32B with Kimi-K2.5 as the only upgrade target; all requests complete, and costs include generation, monitoring, and restarts.
At the largest per-request limit, correction uses 32.8\% less computation than upgrading after completion for similar satisfaction (0.763 versus 0.781); at 20 extra TFLOPs per request, it gains 4.7 percentage points against 2.4.

\section{End-to-End Results}
\label{app:end-to-end}

\paragraph{Second judge.}
GLM-5.3, from a different model family than the judge and every candidate, rescores all 900 evaluation answers with the same two-stage protocol, reasoning enabled, and temperature 0.
Its scores correlate with the original ones at Pearson 0.858 and Spearman 0.840; 95.1\% of the 608 model pairs that both judges strictly order have the same order.
Routing decisions are unchanged, as no method sees evaluation labels; \method reaches 0.5633 against 0.5086 for paced CARROT and 0.4767 for CARROT (\cref{fig:second-judge}), and leads every baseline at all 21 budgets on each trace.

\paragraph{Equal computation.}
Under the same budgets, CARROT spends less in total, running out early in busy windows yet leaving budget idle in others, so \method uses more computation.
We therefore credit each baseline with its best satisfaction among its 21 budget runs that use no more computation per arrival than \method, interpolated linearly; this favors the baselines, which would need a looser budget to spend that much.
\method is then 10.78\% above CARROT, 19.57\% above MixLLM, and at least 46.70\% above the others, and it reaches CARROT's best satisfaction (0.5916 at 45.30~TFLOPs per arrival) with 25.89~TFLOPs, 43\% less.

\paragraph{CARROT with online pacing.}
We keep CARROT's heads, rule, and calibrated starting penalty and, after each request charged computation $c$, update the penalty by multiplicative dual mirror descent~\citep{balseiro2023dualmirror}, $\lambda\leftarrow\lambda\exp(\eta(c-\rho)/\rho)$, clipped to $[10^{-6},10^{6}]$.
The target $\rho$ is the free window budget divided by the expected number of requests left, from the observed arrival rate, the same causal information as our solver.
This pacing is our extension; CARROT itself fixes its penalty in advance, as in the main comparison.
With a fixed penalty, our executor reproduces all 1,470 recorded CARROT runs exactly: the 60 trace streams and ten stationary Poisson streams, each at 21 budgets.
We report the best of nine step sizes from 0.001 to 1, $\eta=0.006$, selected by its satisfaction on the evaluation streams, which no deployed router can do; paced CARROT is thus an optimistic reference rather than an existing baseline.
\method leads it by at least 3.06\% at every budget on either trace (\cref{tab:paced-carrot}) and reaches its best satisfaction, 0.6310 at 12.44~PFLOPs per stream, with 8.20~PFLOPs, a 34\% smaller budget.

\begin{table}[t]
\centering
\small
\setlength{\tabcolsep}{3pt}
\begin{tabular}{llrrr}
\toprule
Trace & Load & Paced CARROT & \method & Gain (\%) \\
\midrule
Azure & $0.5\times$ & 0.6119 & 0.6695 & 9.42 \\
Azure & $1\times$ & 0.5353 & 0.5829 & 8.89 \\
Azure & $2\times$ & 0.4222 & 0.4715 & 11.69 \\
BurstGPT & $0.5\times$ & 0.6261 & 0.6619 & 5.72 \\
BurstGPT & $1\times$ & 0.4968 & 0.5605 & 12.84 \\
BurstGPT & $2\times$ & 0.3767 & 0.4568 & 21.26 \\
\midrule
\multicolumn{2}{l}{Two traces} & 0.5115 & 0.5672 & 10.89 \\
\bottomrule
\end{tabular}
\caption{CARROT with online pacing ($\eta=0.006$) versus \method: mean satisfaction over 21 budgets and ten seeds per load.}
\label{tab:paced-carrot}
\end{table}

\paragraph{Request outcomes.}
Only 2.92\% of \method's arrivals cannot start, against 18.92\% for CARROT, because the solver keeps budget for requests it expects later (\cref{tab:unfinished}).
CARROT's unstarted arrivals come from exhausted window budgets rather than a missing fallback: restricting it to models that the free window budget can start changes its satisfaction only from 0.4778 to 0.4784.
The 4.67\% that \method drops are long and expensive: 82.5\% run on Kimi-K2.5, and finishing one would take 6.8 times the computation of a completed request.
The corrector stops them after a median of 32 tokens, and the released budget serves other requests.

\begin{table}[t]
\centering
\small
\setlength{\tabcolsep}{3.5pt}
\begin{tabular}{lrrrr}
\toprule
Method & Done & Not started & Truncated & Dropped \\
\midrule
\method & 91.88 & 2.92 & 0.52 & 4.67 \\
CARROT & 80.13 & 18.92 & 0.96 & -- \\
RouteLLM & 76.57 & 22.12 & 1.31 & -- \\
MixLLM & 74.57 & 24.11 & 1.32 & -- \\
Unified Cascade & 67.33 & 30.95 & 1.73 & -- \\
RouterBench MLP & 67.24 & 30.99 & 1.77 & -- \\
RouterBench kNN & 66.10 & 32.04 & 1.87 & -- \\
FrugalGPT & 65.29 & 33.14 & 1.58 & -- \\
Unified Routing & 63.92 & 34.16 & 1.92 & -- \\
\bottomrule
\end{tabular}
\caption{Outcome of all arrivals (\%) on the two traces, averaged over 21 budgets, three loads, and ten seeds.}
\label{tab:unfinished}
\end{table}

\section{Ablations}
\label{app:ablation}

\paragraph{Predictor.}
We replace the relative demand predictor with the absolute-satisfaction heads of RouterBench kNN (MiniLM-L12-v2, five cosine neighbors), RouterBench MLP (one 100-unit hidden layer per model), and MixLLM (random forests on BERT-base features), each charged its own encoder and head; the computation predictor, solver, and corrector stay unchanged.
The 13 code-related requests of \cref{fig:ablation}a are selected by a text rule blind to outcomes: 128--511 input characters and a whole-word match for \texttt{code}, \texttt{python}, \texttt{function}, \texttt{program}, \texttt{sql}, \texttt{javascript}, or \texttt{algorithm}.
On 20 Poisson streams of 180 expected arrivals under 9.41~PFLOPs, satisfaction is 0.3232 against 0.2846, 0.2813, and 0.2691; the 95\% interval of the 13.54\% gain is 11.20--15.59\%.

\paragraph{Coordinated solver.}
The fixed-price variant computes, at start-up, the price per FLOP at which training forecasts spend 1/180 of the stream budget per request, and selects $\arg\max_m d_{i,m}-\lambda(\widehat c_{i,m}-\min_{m'}\widehat c_{i,m'})$ among affordable models.
With correction, the coordinated solver is 10.88\% higher, by at least 4.05\% at every budget, and completes 91.88\% of arrivals against 74.83\% at similar computation (26.78 versus 26.14~TFLOPs per arrival); with a fixed price, early arrivals take budget that later ones need, and correction must drop 10.23\% of arrivals to release it.
With the coordinated solver, correction raises satisfaction by 3.72\% on average and at 20 of 21 budgets, by up to 6.54\% at intermediate budgets.

\paragraph{Online allocators.}
Without correction, we compare the solver with two allocators that see the same causal information.
Dual mirror descent pacing uses the update of \cref{app:end-to-end} with its best step size, $\eta=0.01$.
The online multiple-choice knapsack threshold of \citet{zhou2008onlineknapsack}, used by PILOT~\citep{panda2025pilot}, admits a model only if its score per unit of computation is at least $\Psi(z)=(Ue/L)^{z}(L/e)$, where $z$ is the spent budget fraction and $[L,U]$ the 5th and 95th percentiles of training ratios; if none qualifies, it takes the best model within the free budget per expected remaining request.
Pacing moves its price by one step after each charge, and the threshold rises only with budget already spent (\cref{tab:allocators}).

\begin{table}[t]
\centering
\small
\setlength{\tabcolsep}{4pt}
\begin{tabular}{llrr}
\toprule
Alternative & Correction & Min gain (\%) & Wins \\
\midrule
Fixed price & Yes & 4.05 & 21/21 \\
Knapsack threshold & No & 11.04 & 21/21 \\
Pacing ($\eta=0.01$) & No & 3.15 & 21/21 \\
\bottomrule
\end{tabular}
\caption{Gain of the coordinated solver over alternative allocators on the two traces: smallest gain at any of 21 budgets and budgets won.}
\label{tab:allocators}
\end{table}

\paragraph{Runtime correction.}
The 34 Evol-Instruct requests of \cref{fig:ablation}b (median Qwen3-32B answer 1,190 tokens, 421 overall) run on 20 Poisson streams of 180 expected arrivals under 2.74~PFLOPs, with prediction and solver fixed.
Post-response variants apply the FrugalGPT scorer or the Unified Cascade gate to the complete answer and may restart once on a larger model, with all computation charged; for each gate we fix the strongest of its 27 training-fitted policies (plus five FrugalGPT thresholds).
Full correction reaches 0.3082, against 0.2494 (Unified Cascade), 0.2339 (FrugalGPT), and 0.2343 (none), completing 85.82\% of arrivals with 2.40~PFLOPs per stream against 62.37\% and 2.70~PFLOPs for Unified Cascade; the 95\% interval of the 23.57\% gain is 16.62--31.05\%.

\section{Overhead}
\label{app:overhead}

\paragraph{Latency.}
Routing is timed on all 300 evaluation requests, three times each after twelve warmup calls, at batch size one on one thread of a Xeon Platinum 8352V CPU; each call recomputes all features, scores, the benefit curve, and the decision.
Runtime components are timed on every distinct prefix visited in the analysis scenarios, with semantic models on one RTX 4080 SUPER GPU (\cref{tab:overhead-latency}).

\begin{table}[t]
  \centering
  \small
  \setlength{\tabcolsep}{4pt}
  \begin{tabular}{lrr}
    \toprule
    Component & Mean (ms) & P95 (ms) \\
    \midrule
    Initial routing & 45.17 & 86.09 \\
    Solver (included above) & 0.92 & 1.43 \\
    Pre-monitor & 0.33 & 1.17 \\
    Drop: semantic models & 52.36 & 166.10 \\
    Drop: CPU prediction head & 4.05 & 5.51 \\
    Upgrade: encoder & 3.78 & 5.81 \\
    Upgrade: CPU prediction heads & 9.25 & 12.21 \\
    \bottomrule
  \end{tabular}
  \caption{Component latency, timed separately.}
  \label{tab:overhead-latency}
\end{table}

\paragraph{Computation.}
Per arrival, routing, pre-monitoring, drop checks, and upgrade checks consume 0.0025, 0.00002, 1.33, and 0.001~TFLOPs, against 27.08 for source generation and 0.88 for restarts, so routing and monitoring take 4.55\% of all computation (4.44\% end to end).
Relative to inspecting at every position, $\gamma=4$ cuts inspections per arrival from 3.68 to 1.73 and raises completion from 88.90\% to 92.22\%, because the saved budget serves other requests (\cref{tab:gamma}).

\begin{table}[t]
  \centering
  \small
  \setlength{\tabcolsep}{4pt}
  \begin{tabular}{lrrrr}
    \toprule
    Threshold & Satisf. & Inspections & Monitoring & Aux.\ (\%) \\
    \midrule
    Every position & 0.5658 & 3.68 & 2.576 & 8.59 \\
    $\gamma=1$ & 0.5741 & 2.93 & 1.867 & 6.32 \\
    $\gamma=2$ & 0.5769 & 2.32 & 1.575 & 5.36 \\
    $\gamma=4$ & 0.5839 & 1.73 & 1.330 & 4.55 \\
    $\gamma=8$ & 0.5868 & 1.00 & 1.004 & 3.41 \\
    $\gamma=16$ & 0.5834 & 0.31 & 0.341 & 1.13 \\
    \bottomrule
  \end{tabular}
  \caption{Pre-monitor threshold on the analysis scenarios: satisfaction, inspections and monitoring TFLOPs per arrival, and the share of routing and monitoring in all computation.}
  \label{tab:gamma}
\end{table}

\begin{figure*}[t]
  \centering
  \includegraphics[width=.95\textwidth]{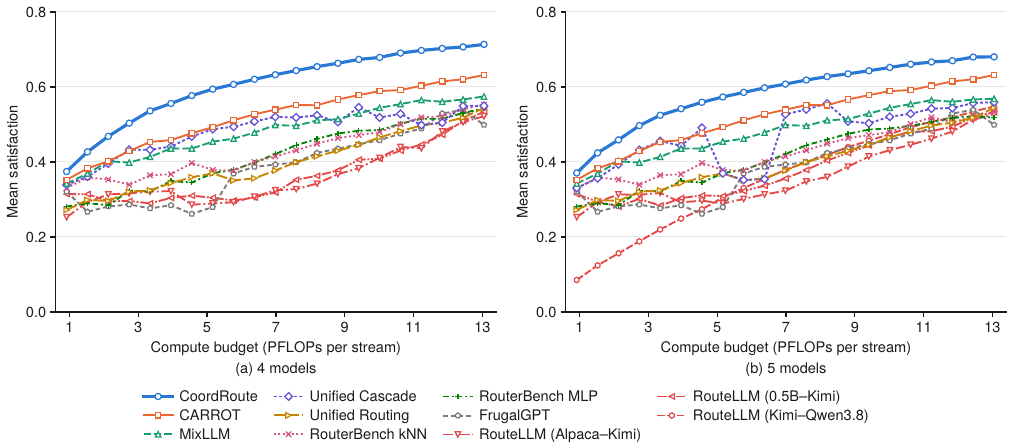}
  \caption{End-to-end comparisons for the expanded pools on the three analysis scenarios. RouteLLM keeps its native two-model form, with model pairs added for each pool.}
  \label{fig:model-pool-baselines}
\end{figure*}

\section{Sensitivity and Robustness}
\label{app:robustness}

\paragraph{Relative model cost.}
The generation cost of Qwen3-32B or Kimi-K2.5 is multiplied by 0.5, 0.75, 1.5, 2, 3, 4, 6, or 8 in prediction, charging, restarts, and every cost comparison, with answers, labels, and streams unchanged; CARROT and MixLLM are recalibrated, and Unified Cascade reruns its search.
At the quartile budgets of the sweep (3.95, 6.98, and 10.02~PFLOP-equivalents per stream), \method improves on CARROT by 10.58--19.82\% averaged over the three budgets and leads CARROT, MixLLM, and Unified Cascade at every multiplier and budget, by 16.95\% on average over the strongest of them.

\paragraph{Model pool.}
The four-model pool adds Qwen2.5-0.5B-Instruct, run locally; the five-model pool also adds Qwen3.8-2.4T-A95B through Alibaba Cloud Bailian, whose reasoning tokens consume budget.
Predictors, calibrations, and policy searches are refit for each pool.
\method leads all baselines at all 21 budgets in both pools (\cref{fig:model-pool-baselines}); the 95\% intervals of its 16.44\% and 12.35\% gains over CARROT are 15.76--17.20\% and 11.51--13.15\%.
Qwen3.8-2.4T-A95B alone reaches 0.8783 but costs 851.94~TFLOPs per answer, ten times Kimi-K2.5, so \method starts only 4.45\% of arrivals on it.

\begin{figure}[t]
  \centering
  \includegraphics[width=\columnwidth]{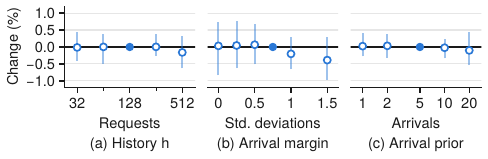}
  \caption{Change in \method's two-trace satisfaction as one solver setting moves from its deployed value (filled): mean (circles) and range (bars) over 21 budgets.}
  \label{fig:solver-settings}
\end{figure}

\paragraph{Ten-times scale.}
Thinning the same intervals with ten times the probabilities gives 1,025--4,394 arrivals per stream; budgets and the reference load are also multiplied by ten.
\method reaches 0.5869 against 0.4858 for CARROT, a 20.81\% gain, completes 94.58\% of arrivals against 81.65\%, and leads at all 21 budgets on both traces (\cref{tab:scale}).
Against CARROT with online pacing ($\eta=0.006$), \method is 11.29\% higher and also leads at all 21 budgets.

\begin{table}[t]
\centering
\small
\setlength{\tabcolsep}{3pt}
\begin{tabular}{@{}lrrrr@{}}
\toprule
Trace & Arrivals & CARROT & \method & Gain (\%) \\
\midrule
Azure & 1,116--4,394 & 0.5002 & 0.5968 & 19.31 \\
BurstGPT & 1,025--4,111 & 0.4713 & 0.5769 & 22.40 \\
\midrule
Two traces & & 0.4858 & 0.5869 & 20.81 \\
\bottomrule
\end{tabular}
\caption{Ten times the arrivals and budget: mean satisfaction over three loads, 21 budgets, and ten seeds per load. Arrivals are mean counts per stream at $0.5\times$ and $2\times$ load.}
\label{tab:scale}
\end{table}

\paragraph{Misestimated reference load.}
All methods receive the same wrong reference load $r\in\{45,90,360,720\}$ instead of 180, and all baselines are recalibrated to it (\cref{tab:reference-load}).
\method changes by at most 0.87\% at any budget because its arrival estimate follows observed arrivals, while calibrated baselines overspend when the load is underestimated (CARROT completes only 44.27\% at $r=45$) and idle when it is overestimated.
Even with an oracle calibration of each static baseline to the actual mean arrival count of each trace and load, which no method knows, \method stays 9.87\% above CARROT and 14.24\% above MixLLM.

\begin{table}[t]
\centering
\small
\setlength{\tabcolsep}{3.5pt}
\begin{tabular}{lrrrrr}
\toprule
Assumed $r$ & 45 & 90 & 180 & 360 & 720 \\
\midrule
\method & 0.5671 & 0.5675 & 0.5672 & 0.5665 & 0.5658 \\
\midrule
CARROT & 66.32 & 45.48 & 18.72 & 20.79 & 41.10 \\
MixLLM & 63.91 & 48.29 & 26.01 & 23.64 & 41.78 \\
RouterBench kNN & 70.71 & 61.72 & 50.56 & 46.96 & 59.35 \\
RouterBench MLP & 75.22 & 65.45 & 52.57 & 54.53 & 67.81 \\
RouteLLM & 70.21 & 61.27 & 59.31 & 70.09 & 74.87 \\
FrugalGPT & 77.64 & 71.84 & 78.89 & 75.96 & 123.17 \\
Unified Routing & 73.85 & 68.26 & 61.24 & 51.81 & 56.69 \\
Unified Cascade & 73.03 & 65.11 & 51.95 & 48.74 & 56.61 \\
\bottomrule
\end{tabular}
\caption{Misestimated reference load on the two traces. First row: satisfaction of \method; other rows: its gain (\%) over each recalibrated baseline.}
\label{tab:reference-load}
\end{table}

\paragraph{Bursts and task-mix drift.}
Synthetic 45-minute Poisson streams place two or three 60--120~s bursts at $k\in\{1,2,4,8,16\}$ times the background rate, with paired layouts and request sequences across $k$ (30 seeds per load; \cref{tab:burst}); from $k=1$ to 16, CARROT's completion falls from 88.9\% to 73.9\% and \method's only from 92.9\% to 88.9\%.
For task-mix drift, each trace arrival is heavy (above the median mean computation) with probability $0.1+0.8\,t/2700$ (rising), $0.9-0.8\,t/2700$ (falling), or 0.66 in busy minutes and 0.16 otherwise (bursty); \method changes by at most 1.9\%.
Solver history, arrival margin, and prior, varied one at a time, keep satisfaction between 0.5650 and 0.5676 (\cref{fig:solver-settings}).

\begin{table}[t]
\centering
\small
\setlength{\tabcolsep}{4pt}
\begin{tabular}{rrrrr}
\toprule
$k$ & \method & CARROT & MixLLM & Paced CARROT \\
\midrule
1 & 0.5992 & 13.95 & 19.89 & 7.62 \\
2 & 0.5973 & 13.76 & 19.46 & 8.05 \\
4 & 0.5886 & 14.29 & 19.72 & 10.04 \\
8 & 0.5638 & 16.66 & 22.90 & 14.83 \\
16 & 0.5288 & 20.73 & 27.91 & 18.50 \\
\bottomrule
\end{tabular}
\caption{Burst intensity: satisfaction of \method and its gain (\%) over the strongest baselines, averaged over three loads, 30 seeds, and 21 budgets.}
\label{tab:burst}
\end{table}

\end{document}